\documentclass[12pt,a4paper]{article}
\usepackage[dvips]{graphicx}
\newcommand{\bea}{\begin{equation}}
\newcommand{\eea}{\end{equation}}
\newcommand{\ber}{\begin{eqnarray}}
\newcommand{\eer}{\end{eqnarray}}

\newcommand{\f}{\frac}

\newcommand{\sig}{\sigma}
\renewcommand{\r}{\right}

\newcommand{\st}{\stackrel{*}}

\begin{document}
\title{Transition to hexagonal pattern under the variation of intrinsic length scales of a reaction diffusion system }
\author{Julian Sienkiewicz \footnote {Email: julas@if.pw.edu.pl} \\ Faculty of Physics\\ Warsaw University of Technology\\ ul. Koszykowa 75\\ 00-662 Warszawa, Poland\\\\ and \\ A. Bhattacharyay\footnote {Email: arijit@fkp.tu-darmstadt.de} \\ Institut f\"ur Festk\"orperphysik,\\ Technische Universit\"at Darmstadt, Hochschulstr.~6,\\ 64289 Darmstadt, Germany\\}

\date{\today}
\maketitle
\begin{abstract}
The intrinsic length scales of a reaction diffusion system (Gierer-Meinhardt model)is varied by quasi-statically changing the diffusion constant of the activator and a transition from rolls to hexagon is detected. The transition is hysteretic or first order like. From stability analysis, we also analytically show the possibility of such transitions.  \\\\
PACS number(s): 87.10.+e, 47.70.Fw
\end{abstract}

\par
Multilateral structures like squares, hexagons, rhombs are observed in chemical patterns \cite{ouyang91}, in Faraday surface wave (\cite{wagner00,arbell02} and references therein),nonlinear optical systems \cite{neubecker03} and in many other systems. Many reaction diffusion systems are studied to show such multilateral structures under various ambient conditions. The origin, stability and dynamics of such multilateral structures are subjects of active investigation \cite{gunaratne94,echebarria00,pena01,golovin03}. In a reaction diffusion system the diffusivities are taken to be constants of activators and inhibitors and the variations in them due to variations in temperature or other experimental conditions are considered to be on very small scales to produce any effect. We argue that if there are some sort of phase transitions which are hysteretic with the variation of diffusion constant, then it indeed has relevance to study transitions under the variation of diffusivities. In a situation when diffusion constants are very close to such transition points, a slight change in them due to the fluctuations in temperature etc. of the system might bring about a phase transition which will not be reversed when the diffusivities are restored. In another respect such studies are worth doing is to understand how such a system behaves under a slow variation of its internal length scales. A slow variation of intrinsic length scale can be effected by slowly varying the diffusion constant of the system. In the present work we are going to represent and characterize the phase transitions brought about by the variation of diffusivity of the activator of Gierer-Meinhardt (GM) \cite{gierer72} model keeping the diffusivity of the inhibitor fixed. In what follows, we will show roll patterns are loosing their stability to hexagonal structures via a first order like transition. Under the action of competiting length scales, the closed multilateral structures characterized by wave numbers which locally satisfy the condition $ q_1 + q_2 + q_3 + ...... = 0 $ are indeed an impressive general outcome and regular hexagons are a more symmetric special case of them.
\par
 The Gierer-Meinhardt model that we have taken up is   \ber\nonumber
\frac{\partial
A}{\partial {t}}&=&D_A{\bigtriangledown} ^2 A +  \rho_A\frac{A^2}{(1+{K_A}A^2)B} - \mu_AA +\sigma_A \\
\nonumber\frac{\partial B}{\partial {t}}&=&D_B{\bigtriangledown}
^2 B +\rho_BA^2 - \mu_BB + \sigma_B \\
\eer 
Where $A$ is the activator concentration and $B$ is that of the inhibitor species. In what follows, we will always take the Turing \cite{turing52} condition $D_B \>> D_A$ between the diffusivities of activator and inhibitor to hold true. In the above expression $\rho_A $ $\&$ $\rho_B$ are the reaction strengths, $\mu_A$ $\&$ $\mu_B$ are the self removal rates and $\sigma_A$ $\&$ $\sigma_B$ are the basic production terms for $A$ and $B$ respectively. In a parameter region where steady spatial order of concentration forms the relaxant length scale is given by \cite{koch94} 
\ber\nonumber
\kappa^2 = \f{C_1}{D_A} - \f{C_2}{D_B}
\eer
where $C_1$ and $C_2$ are constants. Thus the diffusivities are the one which selects the internal length scale of the system. The present report comprises of three parts. The next part shows the results of numerical simulation done under quasi-static variation of the diffusion constant $D_A$. The following part accounts for the observations of numerical simulation on the basis of a stability analysis. The last part includes discussion.
\par
In our simulation we have set $\sigma_A = \sigma_B = 0$. The other parameters are $\rho_A = \mu_A = 0.01$, $\rho_B = \mu_B = 0.02$, $K_A = .25$, $D_B=0.2$ and $D_A$ is varied from $0$ to $0.01$ by the steps of $0.00001$ near the transition regions and by the steps of $0.0001$ in other parts keeping $10^4$ temporal steps between every variation of $D_A$. The simulation is done on two dimensions by implementing finite difference method and on a lattices of size 256 $\times$ 256. In the first part of our two dimensional simulation so long as $D_A$ remains less than approx. $0.004$, the locally isolated concentration peaks show up. Such structures appear for very small value of $D_A$ compared to $D_B$ because of the fact that the activator finds no time to spread compared to very fast inhibitor spreading. As $D_A$ is increased, near $D_A = 0.004$ steady roll structures which are locally parallel appear and persist upto $D_A < 0.008$. As Fig.1(a) shows, many small domains of parallel rolls are present and are oriented in all possible directions. These domains are separated by very many defects and domain walls. At about $D_A= 0.008$ a transition from rolls to almost regular looking hexagonal structures takes place. This phase is stable upto a $D_A$ value about $0.009$. On a further increase in $D_A$, hexagonal patterns loose its stability to a steady homogeneous state. Fig.1(a) and 1(b) show the roll structure at $ D_A = 0.005$ while increasing $D_A$ and hexagonal at $ D_A = 0.0085$ as seen in the lattice. After the hexagons have stabilized at $D_A > 0.008$, now if we go on decreasing $D_A$ the rolls do not come back at $D_A = 0.008$ but at a much lower value. The transition from rolls to hexagons show hysteresis. A plot of average concentration of $A$ against $D_A$ is shown in Fig.2. These plots clearly show a jump in the average concentration as we go from below (circles) along $D_A$ axis which marks the transition to hexagons from rolls at $D_A=0.008$. As we come down the $D_A$ axis (triangles), we see that the average amplitude does not immediately jump back to the lower line at $D_A=0.008$ but continues to follow the same upper line representing the hexagons and comes back to the lower value at around $D_A = .0055$ which marks transition to rolls from hexagons. Thus Fig.2 shows hysteresis and identifies the transition to be first order like. The average activator amplitude also comes out as a relevant parameter to characterize such situations.  
\par
Having numerically found the results let us try to explain it from analytics.
After rescaling of concentration, length and time as $A=\mu_AA$, $l=\sqrt{\f{D_B}{\mu_A}}l$ and $t=\mu_A t$ respectively, Eq.1 takes the form            
\ber\nonumber
\frac{\partial
A}{\partial {t}}&=&\bar{D}{\bigtriangledown} ^2 A +  \bar{\rho_A}\frac{A^2}{(1+\bar{K}A^2)B} - A \\
\nonumber\frac{\partial B}{\partial {t}}&=&{\bigtriangledown}
^2 B +\bar{\rho_B}A^2 - \bar{\mu_B}B \\
\eer
where we have taken $\sigma_A=\sig_B=0$. Under such scaling, $\bar{D}=\f{D_A}{D_B}<1$, $\bar{\rho_A}=\f{\rho_A}{\mu_A^2}$, $\bar{K}=\f{K_A}{\mu_A^2}$, $\bar{\rho_B}=\f{\rho_B}{\mu_A^3}$, and $\bar{\mu_B}=\f{\mu_B}{\mu_A}$. The homogeneous steady fixed point of the system is $A=\st{A}=.0085$ (under prevalent conditions of the simulation) and $B=\st{B}=\f{\bar\rho_B\st{A}^2}{\bar\mu_B}= 85.0$. A linear stability analysis shows that inhomogeneous perturbations will grow with a growth rate $\lambda $ given by
\ber\nonumber
\lambda = &-& \f{1}{2}\left[\bar D k^2 + \f{2\bar K \st{A}^3\bar \rho_B}{\bar \rho_A \bar \mu_B}-1+k^2+\bar \mu_B]\r]\\ & &{\pm} \left[ \left \{(\bar D k^2 + \f{2\bar K \st{A}^3\bar \rho_B}{\bar \rho_A \bar \mu_B}-1)-(k^2+\bar \mu_B)\r\}^2-8\bar \mu_B\r]^{1/2}
\eer

In the range of given parameter values the real part is always negative giving no scope for Hopf modes to appear and in our simulation we have not seen any too. For a roll solution to grow the discriminant has to be greater than zero and its absolute value has to be more than the first part in the expression of $\lambda$. These two conditions give the following inequalities for the range of value of the selected wave number with the variation of $D_A$
\bea\nonumber
k^2 \ge \f{(\surd{2} - \surd{\bar \mu_B})^2 - \f{2\bar K \st{A}^3\bar \rho_B}{\bar \rho_A \bar \mu_B}-1}{\bar D -1}
\eea
and 
\bea\nonumber
k^2 \le 1-\f{2\bar K \st{A}^3\bar \rho_B}{\bar \rho_A \bar \mu_B}
\eea
  
These graphs are plotted on a $k - D$ plane as shown in Fig. 3. In this figure we can easily identify the region where rolls will grow as that bounded by the two curves and the $k$ axis.  In the above expression $k$ is the magnitude of $\bf k$. In our simulation this region has been captured for $D_A>0.004$ to $0.008$. 

The interesting things show up if we try to see how a perturbation grows which tend to rotate the existing roll solutions. To do so we perturb the roll solutions as  $ A(1+\delta a\cos{(\bf k_1.r)})\cos{(\bf k.r)}$ so as to probe a possible rotation of the locally parallel rolls and we end up with two sets of linear decoupled equations in $\delta a$ and $\delta b$ obtained from harmonic balance of the terms containing $\cos {\bf(k_1+k).r}$ and $\cos {\bf(k_1-k).r}$. The equations are
\ber\nonumber
\left(\bar{K}+\f{3A^2}{4}\r)\frac{\partial
\delta a}{\partial {t}}&=&-\left[\bar{D}\left(\bar{K}+\f{A^2}{2}\r){\vert{\bf k_1+k}\vert}^2 + \bar{D}{\vert{\bf k_1-k}\vert}^2\f{A^2}{4}\r ]\delta a + \f{A\bar{K}}{B}\delta a \\\nonumber &-& \left(\bar{K}+\f{3A^2}{4}\r)\delta a \\\nonumber
\frac{\partial
\delta b}{\partial {t}}&=& -{\vert{\bf k_1+k}\vert}^2\delta b -\mu \delta b\\
\eer and
\ber\nonumber
\left(\bar{K}+\f{3A^2}{4}\r)\frac{\partial
\delta a}{\partial {t}}&=&-\left[\bar{D}\left(\bar{K}+\f{A^2}{2}\r){\vert{\bf k_1-k}\vert}^2 + \bar{D}{\vert{\bf k_1+k}\vert}^2\f{A^2}{4}\r ]\delta a + \f{A\bar{K}}{B}\delta a \\\nonumber &-& \left(\bar{K}+\f{3A^2}{4}\r)\delta a \\\nonumber
\frac{\partial
\delta b}{\partial {t}}&=& -{\vert{\bf k_1-k}\vert}^2\delta b -\mu \delta b\\
\eer
   \par
It is apparent from the shape of above two sets of equations that a simultaneous validity of them requires ${\bf k_1}$ to have two values such as ${\bf k_1} = {\bf k_{11}}$ and  ${\bf k_1} = {\bf k_{12}}$  where $\vert{\bf k_{11}}\vert=\vert{\bf k_{12}}\vert$ and ${\bf k_{11}}$ \& ${\bf k_{12}}$ are at angles $\theta$ and $\pi \pm \theta$ with the direction of {\bf k}. Such a choice of $k_1$ predicts development of rhombic structures at the cost of rolls. But we have not been able to stabilize such structures. A regular hexagon grows when $\vert{\bf k_{11}}\vert=\vert{\bf k_{12}}\vert=\vert{\bf k}\vert$ along with $\vert{\bf k_{1}}\vert = 0$ and $\theta = 60$°. Because in such a situation we get all the three wave numbers, one being that of the preexisting roll whose growth rate is the same as other two new modes which are locally rotated at 60° with it, grow equally in the concentration of the activator $A$ obeying the equation as follows. 
\ber\nonumber
\left(\bar{K}+\f{3A^2}{4}\r)\frac{\partial
\delta a}{\partial {t}}&=&-\bar{D}\vert{\bf k}\vert\left(\bar{K}+\f{3A^2}{4}\r)\delta a + \f{A\bar{K}}{B}\delta a - \left(\bar{K}+\f{3A^2}{4}\r)\delta a 
\eer
The growth rate is $\exp\left[-(\bar{D}\vert{\bf k}\vert^2+1)+\f{A \bar{K}}{B(\bar{K}+\f{3A^2}{4})}\r]$. Now putting the relative values of the eigen values corresponding to $A$ and $B$ we get an inequality specifying the wave number arne for the growth of hexagons as
\bea\nonumber
k^2 \le \f{-\f{1}{2}\f{2\bar K \st{A}^3\bar \rho_B}{\bar \rho_A \bar \mu_B} \pm \f{1}{2}\left[\left(\f{2\bar K \st{A}^3\bar \rho_B}{\bar \rho_A \bar \mu_B}-2\r)^2- \f{4\bar K\bar \mu_B}{\bar\rho_B\st{A}(\bar K+3/4)}\r]^{1/2}}{\bar D}
\eea
This curve is also plotted on Fig.3 where it almost superposes on the graph given by the Eq.(5) and hexagonal order should show up in the region below this curve. Thus the hexagons and rolls can coexist over a considerable region of this phase space. In fact in almost all the regions where rolls are stable hexagons are also stable. This fact might account for the fact of obtaining so many small domains of locally parallel rolls and the abundance of defects. As we go on increasing $D_A$, we eventually come to a region where the Eq.(4) is violated and thus rolls become unstable to hexagons for which we are still in the stable region. It is important to note that Fig. 1(b) shows a field of hexagons with very few defects compared to those in Fig. 1(a). This is because the hexagons are formed in a region where rolls are not stable. A further increase in $D_a$ takes us beyond the boundary given by Eq.(8) and hexagons finally loose stability to a homogeneous steady state. With this phase diagram we cannot directly account for hysteresis at the transition region of rolls and hexagons but a wide region of overlap is quite indicative of the fact that such things can happen.         

\par
In the conclusion, we would like to mention a transition from rolls to hexagons has been observed with increasing diffusivity of the activator species in the GM model. The transition region shows hysteresis. Such hysteresis can cause a permanent change in the phase of the system even with a small fluctuation in the diffusivities which are generally treated as constants. We also analytically captured the fact that closed multilateral structures can not simultaneously show up in the concentrations of activator and inhibitor and the system decouples in such a situation. At a very high diffusivity of the activators the hexagons looses stability to homogeneous steady state. We argue that a breakdown of Turing condition might be the cause of such a phase change. 
\section{Acknowledgment}
 J.S. would like to express his gratitude to the organizers of the `` German-Polish-Lithunian Dialogue 2003'' (Especially the hosting group of B. Drossel). He also acknowledges the financial support of the Warsaw University of Technology.
\newpage
{\bf FIGURE CAPTIONS}\\
Figure 1(a).- The figure shows many domains of locally parallel rolls separated by domain walls and other defects at a value of the diffusivity $D_A =0.005$. In this figure the concentration of the activator ($A$) has been plotted on a 256$\times$256 lattice space.\\\vspace{1 cm}

Figure 1(b).- The figure shows regular hexagonal order in the concentration of the activator species $A$ at a value of the diffusivity $D_A = 0.0085$ on the same space lattice as Fig. 1(a).\\\vspace{1 cm}
 
Figure 2.- The figure shows a plot of the average activator concentration against $D_A$ where the circles are plotted for increasing $D_A$ and the triangles are plotted when $D_A$ is decreasing. At $D_A \equiv 0.008$, a jump in the curve indicates transition to hexagons from rolls and at $D_A \equiv 0.0055$, the fall in the curve marks coming back from hexagons to rolls.\\\vspace{1 cm}

Figure 3.- The figure shows the phase diagram as obtained from linear stability on a $D - k$ plane. The roll pattern is stable in the region above the curve '1' and below the curve '2'. The hexagons are stable anywhere below the curve '3'. Thus there is a region of coexistence for rolls and hexagons which lie above curve '2' and below curve'3'. In the region below curve '2' and at the same time below curve '3' only hexagons are stable.

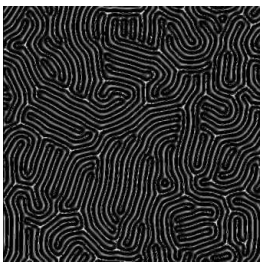
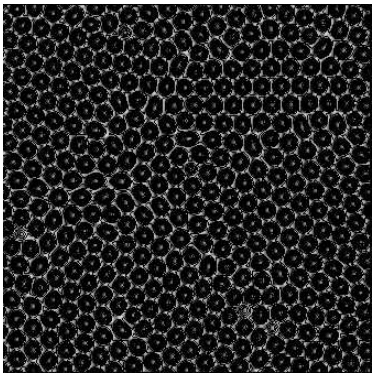
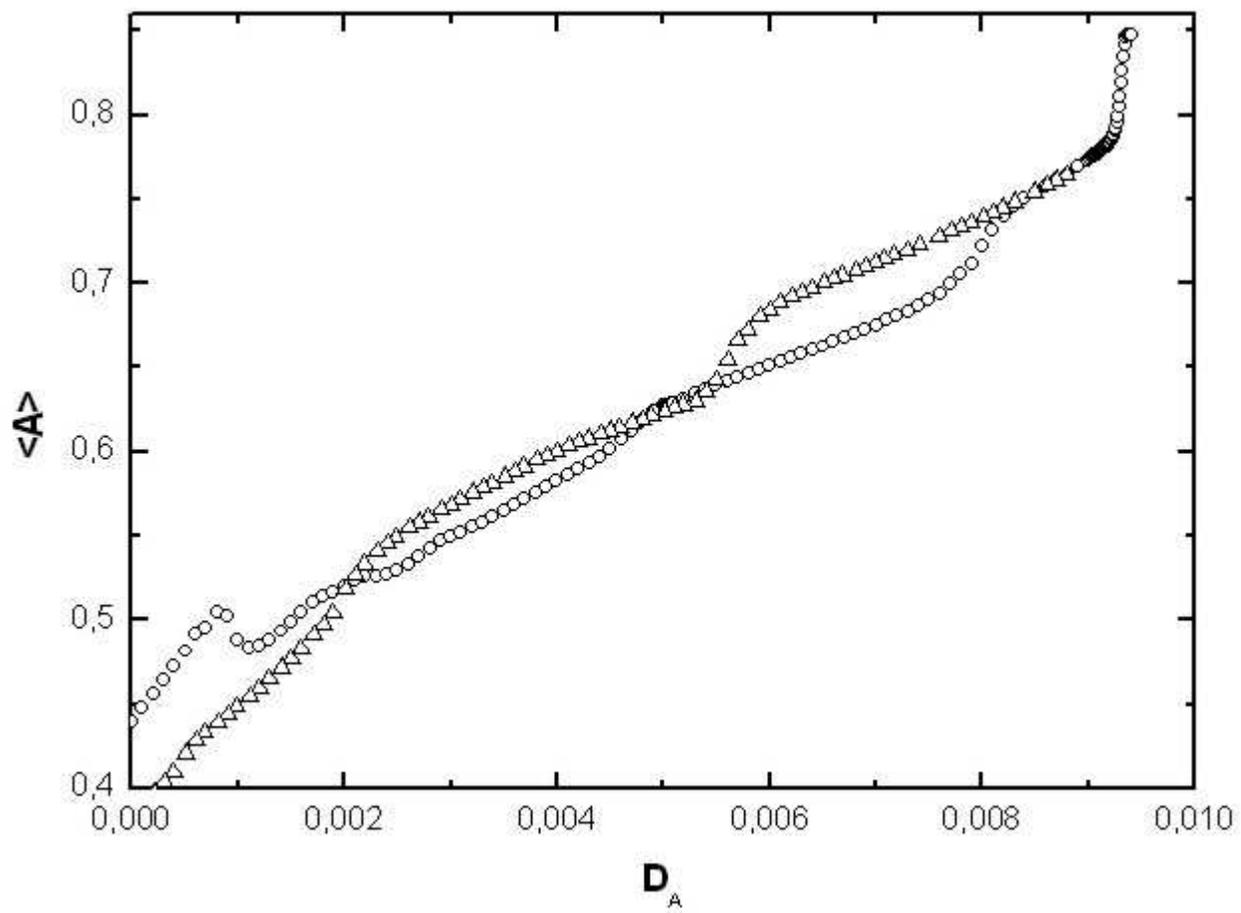
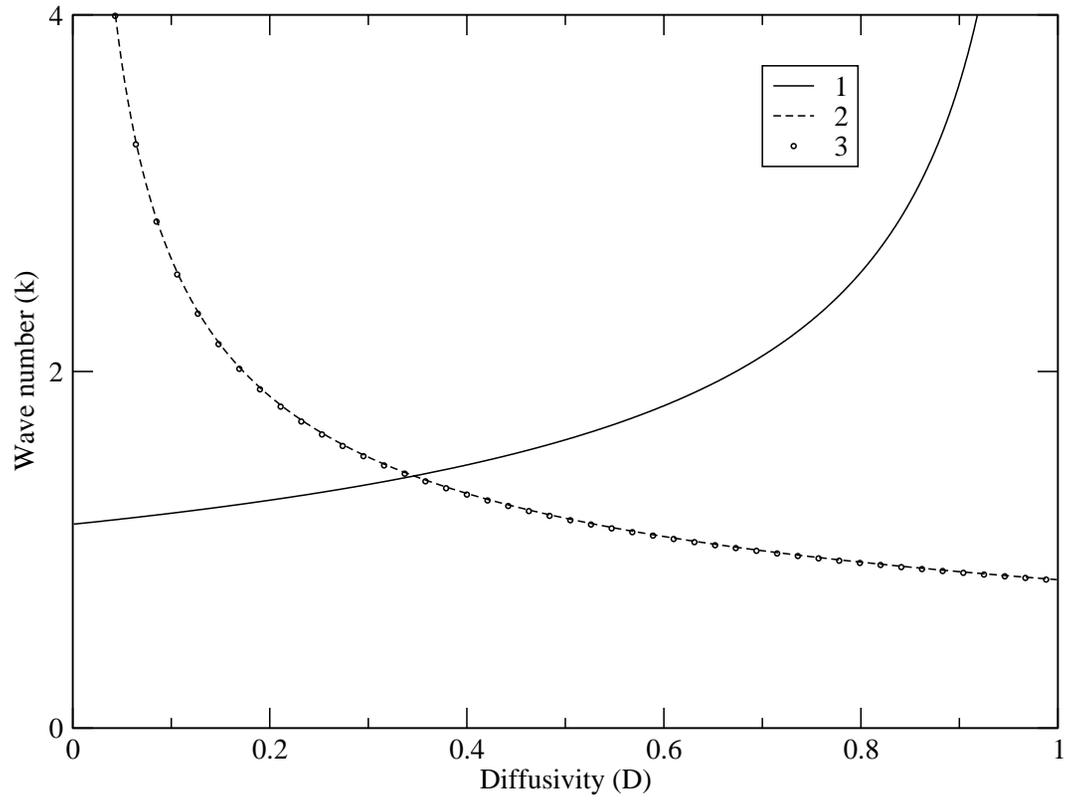
\end{document}